\documentclass[12pt,preprint]{aastex}

\slugcomment{To appear in {\it Spitzer} First Results ApJ Supplement}
\shorttitle{IRAC/NIR imaging of NGC 7129}
\shortauthors{Gutermuth et al.}

\begin{document}


\title{The NGC 7129 Young Stellar Cluster: A Combined {\it Spitzer}, MMT,
and 2MASS Census of Disks, Protostars, and Outflows}


\author{Robert A. Gutermuth\altaffilmark{1}, S. Thomas Megeath\altaffilmark{2}, James Muzerolle\altaffilmark{3}, Lori E. Allen\altaffilmark{2}, Judith L. Pipher\altaffilmark{1}, Philip C. Myers\altaffilmark{2}, \& Giovanni G. Fazio\altaffilmark{2}}


\altaffiltext{1}{Department of Physics and Astronomy, University of Rochester, Rochester, NY 14627 (rguter@astro.pas.rochester.edu)}
\altaffiltext{2}{Harvard-Smithsonian Center for Astrophysics, Mail Stop 42, 60 Garden Street, Cambridge, MA 02138}
\altaffiltext{3}{Steward Observatory, University of Arizona, 933 N. Cherry Ave. Tucson, AZ 85721}


\begin{abstract}

We present the analysis of seven band (1.2 to 8~$\mu$m) ground and space-based 
imaging of the NGC 7129 young stellar cluster from FLAMINGOS on MMT, 2MASS, and 
the Infrared Array Camera (IRAC) on the {\it Spitzer Space Telescope}.
An analysis of the $H-[4.5]$~vs.~$J-H$ colors reveals 84 objects with 
circumstellar disks. Of these, 42 are located within the cluster core, 
a 0.5~pc (100$^{\prime\prime}$) 
radius region of enhanced stellar surface density.  
From a luminosity and extinction limited sample of the stars within the 
cluster core boundary we have determined that 54\%~$\pm$~14\% have 
circumstellar disks.  Finally, we report the detection of several 
resolved outflows in the IRAC 4.5~$\mu$m mosaic.

\end{abstract}


%% subject header.  Objects should be in the appropriate "individual"
%% headers (e.g. quasars: individual, stars: individual, etc.) with the
%% additional provision that the total number of headers, including each
%% individual object, not exceed six.  The \objectname{} macro, and its
%% alias \object{}, is used to mark each object.  The macro takes the object
%% name as its primary argument.  This name will appear in the paper
%% and serve as the link's anchor in the electronic edition if the name
%% is recognized by the data centers.  The macro also takes an optional
%% argument in parentheses in cases where the data center identification
%% differs from what is to be printed in the paper.

\keywords{pre-main sequence --- stars: formation --- infrared:stars}


\section{Introduction}

Excess emission at infrared wavelengths has been frequently used to 
identify those stars that have circumstellar disks.  
\citet{lada00} have shown that disk surveys that utilize 1-2.5~$\mu$m photometry
exclusively 
are usually incomplete, and only by inclusion of longer wavelength photometry 
can a more complete disk sample be achieved.  Such samples, once obtained, 
allow unbiased measurements of the fraction of sources with circumstellar 
disks in a young stellar cluster.  Combined with spectroscopically-derived 
stellar ages, cluster disk fractions may provide insight into the average 
disk lifetime \citep{hll01b}.  Surveying multiple and more distant 
clusters with ground-based $L$--band imaging is impractical however, 
as atmospheric emission severely limits sensitivity.  

Using the Infrared Array Camera (IRAC) onboard the {\it Spitzer Space 
Telescope} \citep{wern04}, we can efficiently observe 
at 3-8~$\mu$m with dramatically improved sensitivity, 
enabling us to obtain a significant census of the YSO population in young 
stellar clusters.  Specifically, the 4.5~$\mu$m channel on IRAC has optimal 
sensitivity to both photospheric and disk emission.  When combined with 
ground-based $JHK$ imaging, these new data allow unbiased measurements of the 
disk fractions of young stellar clusters at distances as large as 1~kpc,
increasing the number of clusters that can be studied this way significantly.
In this paper, we present just such a young stellar object (YSO) census for 
the star forming region 
NGC~7129 to demonstrate the capabilities of IRAC/{\it Spitzer}.

Located at a distance of 1~kpc\footnote{The adopted distance 
of 1~kpc is derived from from high resolution spectroscopy and optical 
photometry of one cluster member \citep{raci68}; other estimates in 
the literature, ranging from 0.9 kpc by \citet{abk00} to 1.25 kpc by 
\citet{sy89}, bracket the 1~kpc value.  The adopted distance does not affect 
our analysis.} 
\citep{raci68}, NGC~7129 is a region 
of bright reflection nebulosity at optical wavelengths, illuminated by 
the Be stars BD+$65^{o}1637$ and BD+$65^{o}1638$.  These two intermediate 
mass stars are part of a cluster of low mass stars \citep{hoda94} which 
occupies a cavity west of a kidney-shaped molecular cloud \citep{ridg03}.  
The cavity is sharply defined in both CO and submillimeter emission 
\citep{font01}, and the cavity wall is 
seen as a nebulous filament detected at both optical and near IR wavelengths.
A third massive member, LkH$\alpha$~234, lies to the east of the main cluster 
on the edge of the cavity.  An optical jet pointing southwest into the cavity 
has been associated with it \citep{ray90} or one of its companions 
\citep{wein96} and may be the counter-jet to a red-shifted
molecular outflow lobe detected to the northeast.  Three arcminutes to the 
south of the cluster lies FIRS2, a deeply embedded intermediate-mass 
protostellar object located at the primary peak in $^{13}$CO emission 
\citep{bech78} and associated with a multipolar molecular outflow 
\citep{fuen01}.  There are several known Herbig-Haro objects in and around the 
region, 
many of which are associated with sites of outflow activity \citep{es83,hl85}.  

\section{Observations}

%% In a manner similar to \objectname authors can provide links to dataset
%% hosted at participating data centers via the \dataset{} command.  The
%% second curly bracket argument is printed in the text while the first
%% parentheses argument serves as the valid data set identifier.  Large
%% lists of data set are best provided in a table (see Table 3 for an example).
%% Valid data set identifiers should be obtained from the data center that
%% is currently hosting the data.

For information on the IRAC instrument, see \citet{fazi04}.
The IRAC/{\it Spitzer} observations were taken as part of the Spitzer young 
stellar cluster survey.  The observations are described in detail 
by \citet{mege04}.

Near~IR observations of the region in the $J$~(1.2~$\mu$m), $H$~(1.6~$\mu$m), 
and $K_{s}$~(2.2~$\mu$m) wavebands were obtained by the authors on 
June~15,~2001 using the FLAMINGOS instrument \citep{elst98} on the 6.5 meter 
MMT Telescope.  A 2~$\times$~2 position raster was used to make a 
$9^{\prime} \times 9^{\prime}$ mosaic.  Five dithered mosaics were 
obtained in each band, for a total exposure time of 75~seconds at $J$, $H$, 
and $K_{s}$.
The data were processed using custom IDL routines developed 
by the authors for ground-based near~IR data reduction which include 
modules for linearization, flat-field creation and application, background 
frame creation and subtraction, distortion measurement and correction, and 
mosaicking.   

Point source detection and aperture photometry of all point sources were 
carried out using PhotVis version 1.09, an IDL GUI-based photometry 
visualization tool developed by Gutermuth.  PhotVis utilizes DAOPHOT modules 
ported to IDL as part of the IDL Astronomy User's Library \citep{land93}.  
Detections were visually inspected and those that were identified as 
structured nebulosity were considered non-stellar and rejected.  Radii of 
the apertures and inner and outer limits of the sky annuli were 
1$^{\prime\prime}$, 2$^{\prime\prime}$, and 3.2$^{\prime\prime}$ respectively 
for the near~IR data and 2.4$^{\prime\prime}$, 2.4$^{\prime\prime}$, 
and 7.2$^{\prime\prime}$ respectively for the IRAC data.  
FLAMINGOS photometry was calibrated by minimizing residuals to corresponding 
2MASS detections, using only those objects with $H-K_{s}$ $<$ 0.6 mag to minimize 
color differences in the 2MASS and FLAMINGOS filter sets.  Photometry for 
stars that were non-linear or 
saturated in the FLAMINGOS data were replaced with the appropriate 2MASS 
measurements in the final analysis.  IRAC photometry was calibrated using 
large aperture measurements of several standard stars from observations 
obtained in flight.  An additional correction was 
derived and applied for each channel to correct for the smaller apertures used 
in this study. 

%% In this section, we use  the \subsection command to set off
%% a subsection.  \footnote is used to insert a footnote to the text.


%% This section also includes several of the displayed math environments
%% mentioned in the Author Guide.

\section{Results\label{results}}

The four-color composite IRAC image is shown in Figure~\ref{image}. 
We note two different types of extended 
emission.  
The brightest and most extended nebulosity is the reflection nebula,
which is most prominent in the 8.0~$\mu$m (red) band.  This band includes
the 7.7~$\mu$m emission feature commonly attributed to PAHs.
In addition, 4.5~$\mu$m-bright (green) structured shocks 
and knots associated with outflow activity are easily discernible at two sites
located south and east of the cavity.  The preponderance of outflow activity 
outside the cluster core suggests that current star formation activity 
is ongoing in the molecular cloud.

Centered in the reflection nebulosity is the cluster of stars most apparent 
in the near~IR \citep{hoda94}.
Using our $K_{s}$--band 
star counts, we determined the location of the peak local stellar surface 
density, and measured the radial stellar surface density profile centered on 
this point.  We measured the half-width at half maximum density from this 
profile to be approximately 0.17~pc, and defined the boundary of the cluster 
core as a circle centered on the stellar density peak with a radius of three 
times this distance, or 0.5~pc \citep{gute04}.  

\subsection{Census of Objects with Disks}

To identify the population of young, forming stars with IR 
excess (circumstellar disks) we use an $H-[4.5]$~vs.~$J-H$ color-color diagram
for all stars detected at $J$, $H$, and $K_{s}$ with photometric uncertainties of 
less than 0.1~magnitudes and 4.5~$\mu$m (Fig.~\ref{all}b) with 
photometric uncertainties of less than 0.25~magnitudes\footnote{Photometric 
uncertainties take into account varying background toward each star.  This 
component can dominate the 4.5~$\mu$m band error estimates.}.  
We interpret those that 
%%Stars that are more than 1~sigma below and to the right of the lower 
%%reddening vector, which originates at the colors of an M5 dwarf star 
%%measured by IRAC and 2MASS (Brian Patten, personal communication), 
exhibit colors inconsistent with reddened dwarf or giant stars   
as stars with circumstellar disks.  
%%We interpret these as stars with circumstellar disks.  
We used the reddening law of \citet{rl85} for this analysis, interpolated to 
take into account the filter response of the 4.5~$\mu$m channel of IRAC.  
%%Note the significantly smaller number of disks in the $JHK$ color-color 
%%diagram for the same objects (Fig.~\ref{all}a).
We have restricted this study to only those objects with dereddened $J$--band 
magnitudes less than 16.5, equivalent to the luminosity of a 1~Myr old 
0.06~M$_{\odot}$ or a 
3~Myr old 0.08~M$_{\odot}$ star \citep{barr98} at the adopted distance.  
An age of 3~Myr was chosen as a conservative upper limit for the age of the 
cluster core population, although 
it is clear from the presence of protostellar objects reported in this 
paper and the companion studies that there are much younger stars 
present in the molecular cloud.

\begin{figure}
\epsscale{1}
\plotone{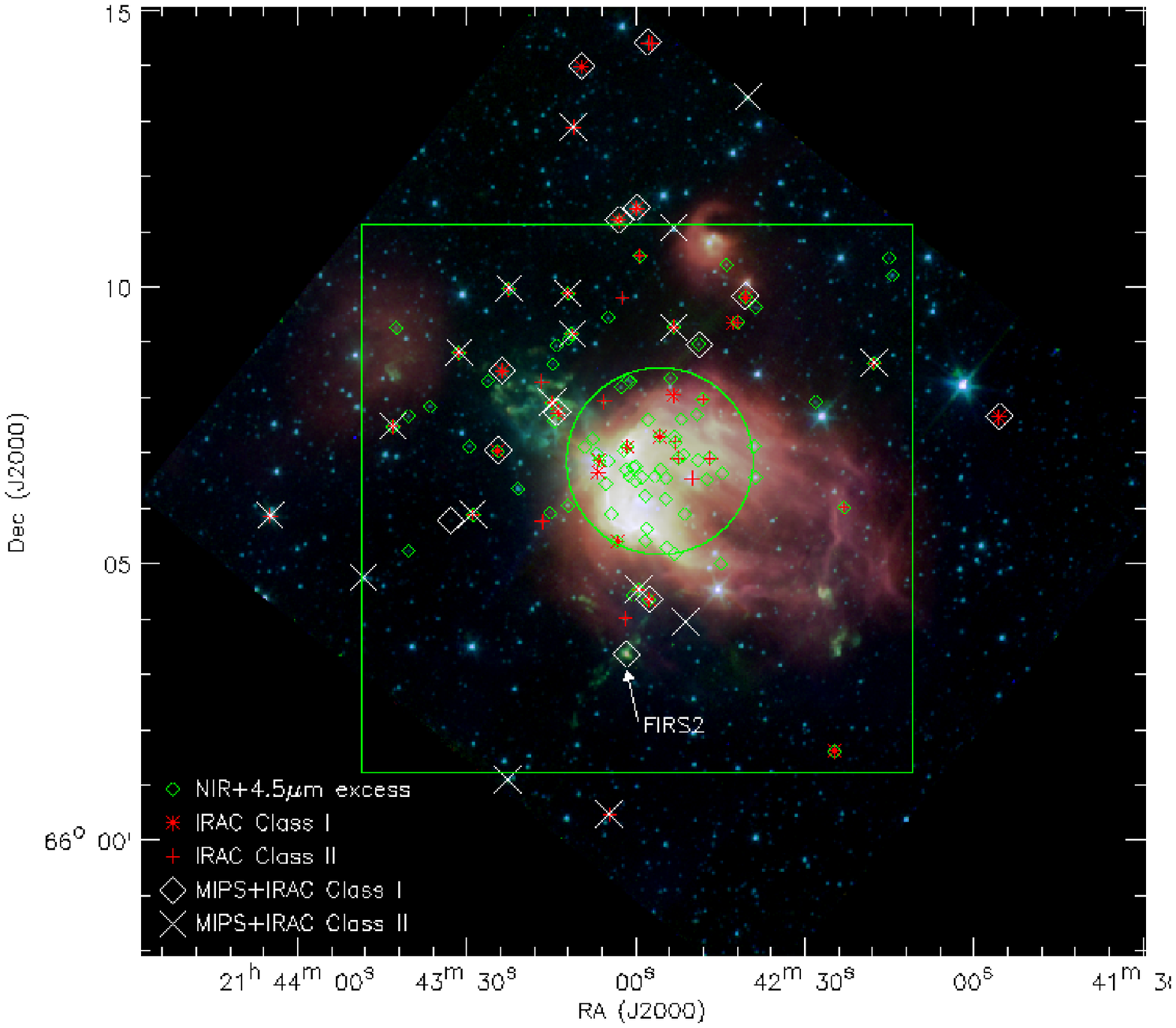}
\caption{IRAC four-color image of NGC~7129.  The large green square shows 
the field of view of the near~IR data.  The green circle denotes the cluster 
core boundary as defined in Section~\ref{results}.  The green diamonds are 
those objects with disks from the $H-[4.5]$~vs.~$J-H$ color-color diagram 
in Fig.~\ref{all}.  Embedded protostars (Class~I YSOs) and 
classical T-Tauri stars (Class~II YSOs) reported in the companion papers are 
also marked.  Red asterisks are Class~I YSOs and red pluses are 
Class~II YSOs as determined in \citet{mege04} using IRAC photometry
alone and YSO modeling from \citet{alle04}.  Large white diamonds are 
Class~I YSOs and large white X's are Class~II YSOs from IRAC 
and MIPS combined photometry in \citet{muze04}.\label{image}}
\end{figure}

The positions of those objects with IR excess are plotted as green diamonds 
in Figure~\ref{image}.  We find 84 objects with disks out of a total of 405 
objects that are detected in all four bands within the above constraints.  
Of these, 42 out of the 87 that are located within the 
cluster core boundary have disks.  Note that the cluster core has 
a significantly higher density of objects with disks compared with the rest 
of the 
field even though the 4.5~$\mu$m sensitivity is significantly reduced because 
of the bright nebulosity in the cavity.  Nevertheless,
the total number of stars with disks detected both inside and 
outside the cluster core boundary is similar.  Most of this peripheral star 
formation appears to be located in parts of the molecular cloud north, 
east, and south of 
LkH$\alpha$~234.  There are also several stars with IR excess near the western 
edge of the reflection nebula.  These may be young stars with disks 
from the core that have emerged from the molecular cloud,
or they could be background galaxies, planetary nebulae, or AGB stars.  
The IRAC 
colors and the number of these potential contaminating objects are 
currently not well characterized, but given the small number of IR excess 
sources detected away from the molecular cloud, these contaminants are not 
expected to alter the deduced disk frequency significantly.  

For comparison, red points are also plotted in Figure~\ref{image} to mark 
those YSOs detected and classified using only the four IRAC bands 
\citep{mege04,alle04}, and white points mark those YSOs detected and 
classified using the IRAC and MIPS combined colors \citep{muze04}.  These 
methods complement each other well, as the longer wavelengths suffer from
loss of sensitivity in the bright central nebulosity, whereas the near IR 
data in this study cannot detect sources in the heavily extinguished 
regions of the molecular cloud.
It is clear in all these studies that in addition to the dense cluster core 
there is an extended population of young stars predominantly located in 
the eastern molecular cloud.  

\subsection{Measuring the Disk Fraction}

\begin{figure}
\epsscale{1}
\plotone{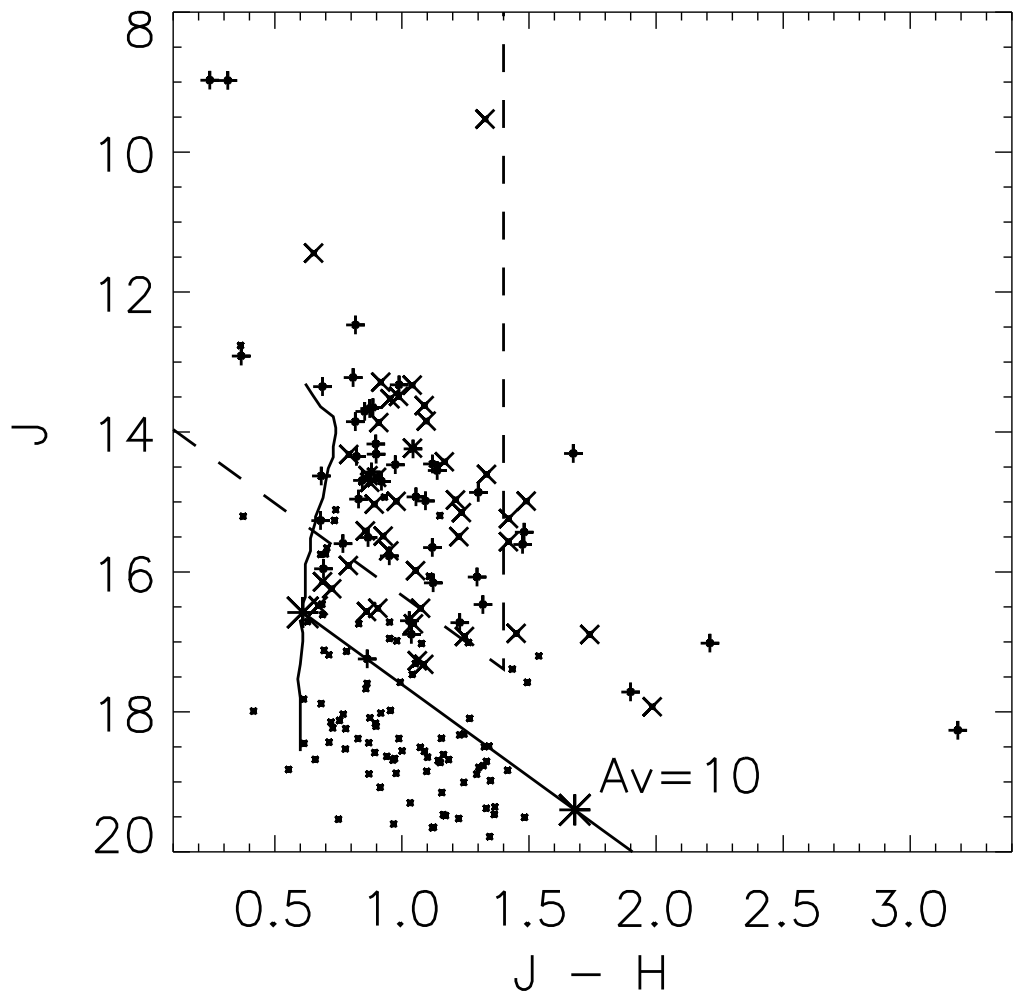}
\caption{$J-H$~vs.~$J$ color-magnitude diagram of the objects in the 
cluster core boundary.  Dots are objects only detected at $J$, $H$, and $K_{s}$ 
but not 4.5~$\mu$m.  Crosses and pluses mark those objects that have been 
detected at $JHK$ and 4.5~$\mu$m and have 4.5~$\mu$m excess emission and no 
excess emission respectively.  For comparison, the near vertical curve is 
the 3~Myr isochrone of 
\citet{barr98} for stars between 1~M$_{\odot}$ and 0.025~M$_{\odot}$.  The 
slanted line is the \citet{rl85} reddening vector for a 0.08~M$_{\odot}$ star.
The slanted dashed line denotes the limiting dereddened $J$ magnitude 
corresponding to 0.2~M$_{\odot}$ and the 
vertical dashed line denotes the limiting extinction of A$_{V} =$~6~mag, 
chosen to give an unbiased sample for measuring the disk 
fraction.\label{frac1}}
\end{figure}

To achieve an unbiased sample of stars in the NGC~7129 cluster for 
%%accurate 
measurement of the fraction of stars with disks, we first
restrict our sample to objects detected at $J$, $H$, and $K_{s}$ wavelengths
within the cluster core boundary.  We then plot our sample on a 
$J-H$~vs.~$J$ color-magnitude diagram (see Fig.~\ref{frac1}), marking the stars 
which are also detected at 4.5~$\mu$m.
%% with photometric uncertainties less than 0.25~magnitudes.  
Since stars with disks are substantially more luminous at 4.5~$\mu$m,
the disk fraction can be overestimated if we select sources on the basis 
of the IRAC photometry.  By basing our sample selection on the 
dereddened $J$--band magnitude, which
is dominated by the emission from the stellar photosphere \citep{kh90}, 
we avoid this bias.   Furthermore, by adopting
a modest maximum extinction, we are sampling all stars above the magnitude
limit within the spatial volume defined by our
cluster core boundary and the extinction limit.
We choose a dereddened $J$ magnitude upper limit of 15.5, corresponding to the 
luminosity of a 1~Myr old 0.1~M$_{\odot}$ star or a 3~Myr old 0.2~M$_{\odot}$ 
star \citep{barr98} at the adopted distance, and an extinction upper limit 
of A$_{V} =$~6~mag (A$_{J} =$~1.7~mag).  These limits are chosen such that 
90\% of the stars in the sample have corresponding 4.5~$\mu$m detections, 
with the non-detections evenly distributed on the color-magnitude diagram.  
%%The final luminosity-limited and extinction-limited sample is 
%%intended to minimize bias in detecting disks by being complete down to 
%%the given dereddened $J$ magnitude 
%%within the cluster boundary.  
%%We note that this sample only 
%%reaches [4.5]~$=$~14~mag because of the reduced sensitivity in the bright 
%%nebulosity coincident with the cavity region.  Using star counts from several
We also take the same sample from nearby control fields on the edge of our 
mosaic.  From this we estimate 20\% of the stars in the final cluster core 
sample
%%with 4.5~$\mu$m magnitudes as faint as 14 to be 
are field stars.

Those stars that remain in our final sample are plotted on the 
$H-[4.5]$~vs.~$J-H$ color-color diagram in Figure~\ref{frac2}.  
All stars to the right and 
below the reddening vector originating at the M5 dwarf colors are 
considered to have excess emission at 4.5~$\mu$m and therefore 
circumstellar disks.  Of the 62 stars in the final sample, 32 have disks.  
If 20\% of the sample of 62 stars are in fact background stars, then only 50 
are probable cluster members, yielding a high estimated disk fraction of 64\%. 
If we reject 
all those objects within 1 sigma of the M5 reddening vector as before, 
the total number of objects with disks drops to 27, for a total of 
54\%~$\pm$~14\%.  The statistical uncertainty is derived assuming Poisson 
counting statistics for the number of stars with excesses, the total number of 
stars in the cluster core, and the number of stars in our control fields used 
for determining background contamination.

\section{Discussion}

\begin{figure}
\epsscale{1}
\plotone{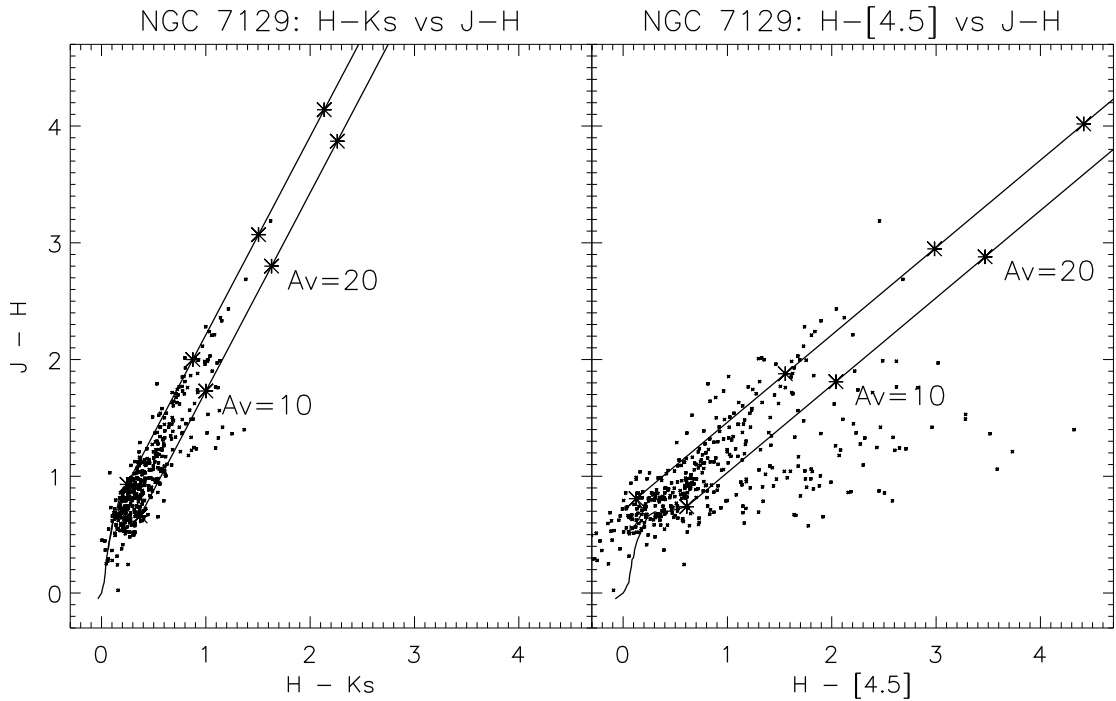}
\caption{$JHK_{s}$ and $JH[4.5]$ color-color diagrams for the NGC~7129 region.  
The upper and lower reddening vectors show the \citet{rl85} reddening law for 
an M5 giant \citep{bb88} and an M5 dwarf respectively, 
as measured by IRAC and 2MASS (Brian Patten, personal communication).  
Those objects greater than 1 sigma below and to the right of 
the lower reddening vector have circumstellar disks.\label{all}}
\end{figure}

\begin{figure}
\epsscale{1}
\plotone{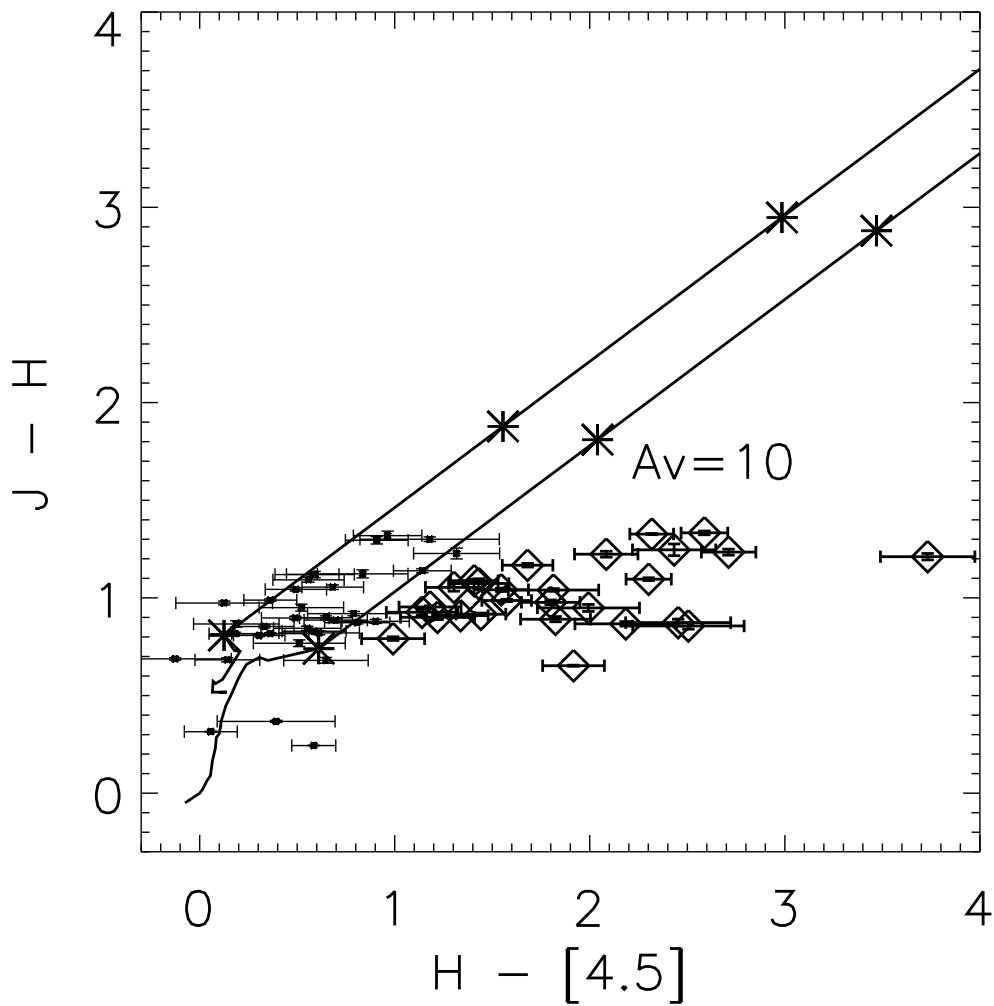}
\caption{$H-[4.5]$~vs.~$J-H$ diagram of the final luminosity and extinction 
limited sample with photometric uncertainties overlaid.  The error bars take 
into account varying sky background toward each star.  This component can 
dominate the uncertainties in the 4.5~$\mu$m band.  Reference colors and 
reddening vectors are the same as those described for Fig.~\ref{all}b.  Those 
objects plotted with diamonds have disks.  From this plot we measure a disk 
fraction of 54\%~$\pm$~14\%.\label{frac2}}
\end{figure}

As demonstrated clearly in Figure~\ref{all}, combining the longer wavelength 
IRAC data with the near~IR data is a more robust method for the detection 
of circumstellar disks than $JHK$-based methods allow.  The photospheres of 
both 
ordinary stars and young stars are characterized well by $J$ and $H$ bands, 
but detecting the warm inner disks at $K$--band is uncertain because of the 
relative 
weakness of disk emission compared to that from photospheres at 2.2~$\mu$m.
The spectral energy distributions of protostellar disks dominate photospheric 
SEDs at 
wavelengths longer than 2.2~$\mu$m, allowing easier detection if adequate 
sensitivity can be achieved, as has been done with IRAC/{\it Spitzer}.  
For disk fraction studies specifically, the 4.5~$\mu$m channel of IRAC has the 
ideal combination of sensitivity to both photospheric and disk emission needed 
for unbiased sampling of the entire stellar population, both with and without 
disks. 

We measure a disk fraction of 54\% for the cluster core of NGC~7129.  When 
compared to other long wavelength ($JHKL$) measurements, we note that 
the disk fraction in NGC~7129 is significantly lower than 
80\% reported for the Trapezium cluster by \citet{lada00} 
and 86\% in NGC 2024 as reported by \citet{hll00}.  Both of these clusters are 
considered to be very young ($<$~1~Myr).  However, our result is fairly close 
to that found for the 2.3~Myr cluster IC~348, which has a disk fraction of 65\% 
\citep{hll01a}, and the 3.2~Myr cluster NGC~2264 which has 52\% \citep{hll01b}.
Future work will 
be focused on measuring the age of the NGC~7129 cluster core to compare with
the disk lifetime study presented in \citet{hll01b}.

While this study concentrated on the cluster core, the significant number of 
young stars and protostellar objects outside the core
%%, and particularly the presence of protostellar objects, 
%%is evidence that star formation is ongoing in the periphery of this region.  
is evidence of active extended star formation in this region. 
Star counts using our $K_{s}$--band photometry
in the cluster core and in our control fields suggest there are approximately 
80 cluster members down to $K_{s} =$~16~mag.  This sensitivity allows us to 
reach unattenuated 1-3~Myr old objects down to 0.055~M$_{\odot}$ \citep{barr98}.
In the periphery of the cluster we estimate the number of stars with disks at 
62, through a combination of those reported here and the YSOs reported in 
the companion papers \citep{mege04,muze04}.  Most of those YSOs that are 
missed in this study are either outside the coverage of the near~IR data or 
are highly embedded and thus missing $J$--band detections.
Clearly, the extended cluster population contains a roughly equal number of 
stars as the high density cluster core, and thus comprises a 
significant portion of the total stellar content in NGC~7129.

The bright nebulosity extending from the northeast boundary of the 
reflection nebula
is coincident with a complex of Herbig-Haro objects, several molecular
hydrogen knots, and a CO outflow; these data indicate that the
nebulosity is probably heated by shocks, and not by radiation from nearby 
stars.  Although the nebulosity is evident in all four IRAC bands, it shows 
a distinctly different color and structure than the reflection nebula, and is 
particularly prominent in the 4.5~$\mu$m band.  Since this band contains the CO 
fundamental and several H$_{2}$ lines, the emission in the 4.5~$\mu$m band 
probably has a large component 
from shock heated molecular gas.  It is still not known whether this region 
contains one or multiple outflows, 
nor is it known which sources are driving the outflow.  IRAC and MIPS
imaging of NGC~7129 have shown four protostellar (Class I) and two
pre-main sequence stars with disks (Class II)
within the nebulosity \citep[this work]{mege04,muze04}.  Given the
complex shape of the emission, and the presence of multiple
protostellar objects, we find it likely
that this nebulosity contains the combined emission from several
outflows.

%%The outflow arcs traced by 4.5~$\mu$m-bright knots detected in the region 
%%south of the reflection nebula are also very interesting.  
Several distinct outflow arcs traced by 4.5~$\mu$m-bright knots appear to be 
associated with FIRS2.  The object itself is elongated perpendicular to the 
extended outflow arcs, suggesting another outflow axis which is resolved as 
distinct emission knots in our $K_{s}$--band image. 
The observed multipolar nature of this outflow system, in general agreement 
with the the outflow analysis of \citet{fuen01}, lends support to the claim 
by \citet{misk01} that FIRS2 is a multiple protostellar system.  We note 
at least one other obvious outflow arc that does not seem to be associated 
directly with FIRS2.  Future work will involve numerical modeling of these 
resolved outflows to investigate their influence on the star forming 
environment of the natal molecular cloud.

\acknowledgments

This work is based on observations made with the {\it Spitzer Space
Telescope}, which is operated by the Jet Propulsion Laboratory,
California Institute of Technology under NASA contract 1407. Support
for this work was provided by NASA through Contract Number 1256790
issued by JPL/Caltech.
Support for the IRAC instrument was provided by NASA through Contract
Number 960541 issued by JPL.
Ground-based observations reported here were obtained at the MMT Observatory, a joint 
facility of the Smithsonian Institution and the University of Arizona.
FLAMINGOS was designed and constructed by the IR instrumentation group
(PI: R. Elston) at the University of Florida, Department of Astronomy
with support from NSF grant (AST97-31180) and Kitt Peak National
Observatory.
This publication makes use of data products from the Two Micron All Sky 
Survey, which is a joint project of the University of Massachusetts and 
the Infrared Processing and Analysis Center/California Institute of 
Technology, funded by the National Aeronautics and Space Administration 
and the National Science Foundation.
This research has made use of the SIMBAD database, operated at CDS, 
Strasbourg, France.
Link to the data set used in this analysis:
\dataset{ads/sa.spitzer\#0003655168}

\clearpage

%% the file, uncomment the \plotone, \plottwo, and
%% If you need a layout that cannot be achieved with \plotone or
%% \plottwo, you can invoke the graphicx package directly with the
%% \includegraphics command or use \plotfiddle. For more information,
%% please see the tutorial on "Using Electronic Art with AASTeX" in the
%% documentation section at the AASTeX Web site,
%% http://www.journals.uchicago.edu/AAS/AASTeX.
%%
%% The examples below also include sample markup for submission of
%% supplemental electronic materials. As always, be sure to check
%% the instructions to authors for the journal you are submitting to
%% for specific submissions guidelines as they vary from

%% This example uses \plotone to include an EPS file scaled to
%% 80% of its natural size with \epsscale. Its caption
%% has been written to indicate that additional figure parts will be
%% available in the electronic journal.

%\onecolumn

%\twocolumn

\end{document}